\documentclass[12pt]{article}
\usepackage{pdproc} 
\usepackage{amsmath}
\usepackage{amssymb}
\usepackage{epsfig}
\usepackage{graphicx}
\usepackage{cite}

\newcommand{\ie}{{\it i.e.}}

\newcommand{\eg}{{\it e.g.}}

\newcommand{\etc}{{\it etc.}}
\newcommand{\eq}{Eq.}

\newcommand{\fig}{Figure}

\newcommand{\Ref}{Ref.}
\newcommand{\Refs}{Refs.}

\newcommand{\Tab}{Table}

\newcommand{\JHFHK}{\mbox{{\sf T2HK}}$^*$}

\newcommand{\stheta}{\sin^22\theta_{13}}
\newcommand{\deltacp}{\delta_\mathrm{CP}}
\newcommand{\ldm}{\Delta m_{31}^2}

\newcommand{\equ}[1]{\eq~(\ref{equ:#1})}
\newcommand{\figu}[1]{\fig~\ref{fig:#1}}

\newcommand{\bi}{\begin{itemize}}
\newcommand{\ei}{\end{itemize}}

\begin{document}

\renewcommand{\thefootnote}{\alph{footnote}}
  
\title{
 Neutrino Factory and Beta Beam Experiment}

\author{Walter Winter}

\address{School of Natural Sciences, Institute for Advanced Study, Princeton, NJ 08540, USA\\
 {\rm E-mail: winter@ias.edu}}

\abstract{I summarize the current understanding 
of the neutrino oscillation physics with a neutrino factory and a beta beam, where I emphasize more recent phenomenological developments.}
   
\normalsize\baselineskip=15pt

\section{Introduction}

Neutrino factories~\cite{Geer:1998iz} could be the ultimate high precision instrument for neutrino
oscillation physics. Compared to superbeams, the neutrinos are produced by
muon decays, which means that the flavor composition is exactly known. 
Neutrino factories are, for small signals, limited by 
neutral currents and the requirement of charge identification in the detector.
Beta beams~\cite{Zucchelli:2002sa} with sufficiently high gamma~\cite{Burguet-Castell:2003vv} (the boost factor determining the neutrino energies) could be an interesting alternative to
neutrino factories, because they have a well-known flavor composition, too, and they do not require charge identification. In this talk, I discuss the neutrino oscillation physics of neutrino factories and beta beams, where I focus on physics rather than machine-related challenges. In addition, note that
neutrino factories and beta beams have applications other than neutrino oscillations,
although neutrino oscillations may be the primary motivation. For example, a neutrino factory
front-end could be used for high statistics tests of rare (flavor-violating) muon decays. Furthermore,
neutrino cross section measurements are an important prerequisite for any future long-baseline 
neutrino oscillation program. As far as the timescale of this talk is concerned, I will primarily focus on neutrino oscillation physics beyond the coming ten years. However, note that beta beams can,
 depending on the purpose, be built on different scales of gamma, which means that (if based on existing equipment) earlier applications could be possible.

\section{Neutrino factory}

The neutrino factory concept (see \Refs~\cite{Geer:1998iz,Albright:2000xi,Apollonio:2002en} and references therein) includes many components. As for superbeams, protons (typically of energies around $8 \, \mathrm{GeV}$) hit a target
to produce pions (and kaons). Compared to superbeams, not the neutrinos from the following pion decays
make up the beam, but the muons from these decays are collected. In order to accelerate the muons further, they need to come in bunches with very little longitudinal and transversal spread, \ie, they need to be ``cooled''. The muons are then accelerated up to typically $20$ to $50 \, \mathrm{GeV}$ and then injected into a storage ring with long straight sections. The neutrino beam is then produced by the decays of the muons in these straight sections.
For example, for muons in the storage ring, we have
\begin{equation}
\mu^- \rightarrow e^- + \bar{\nu}_e + \nu_\mu \, , 
\end{equation}
\ie, equal amounts of electron antineutrinos and muon neutrinos are produced. For three-flavor effects, the most relevant oscillation channel is $\bar{\nu}_e \rightarrow \bar{\nu}_\mu$
(``golden appearance channel'')~\cite{DeRujula:1998hd,Barger:1999jj,Cervera:2000kp}. Obviously, the $\bar{\nu}_e$ oscillating into $\bar{\nu}_\mu$ and producing $\mu^+$ in the detector have to be distinguished from the $\nu_\mu$ staying $\nu_\mu$ (``disappearance channel'') and producing $\mu^-$ (``wrong-sign muons'') in the detector. Therefore, charge identification 
is a key ingredient for a neutrino factory detector. Other technological challenges are rather
large proposed target powers, the muon cooling, and possibly steep decay tunnels.

As far as the ``typical'' parameters for a neutrino factory are concerned, the goal is to achieve about $10^{21}$ useful muon decays per year. There are different approaches for that, such as 
a triangular-shaped storage ring operated with muons and antimuons successively, and a
racetrack-shaped storage ring operated with muons and antimuons (circulating in different directions) simultaneously.
Other ``typical'' parameters are $E_\mu \simeq 20 - 50 \, \mathrm{GeV}$ and $L \simeq 2 \, 000 - 4 \, 000 \, \mathrm{km}$ for leptonic CP violation, as well as magnetized iron detectors with a fiducial mass between $50$ and $100 \, \mathrm{kt}$ (see, \eg, \Ref~\cite{ISS}).

\subsection{Parameter extraction: Correlation and degeneracy problems}

\begin{figure}
\begin{center}
\includegraphics[width=8cm]{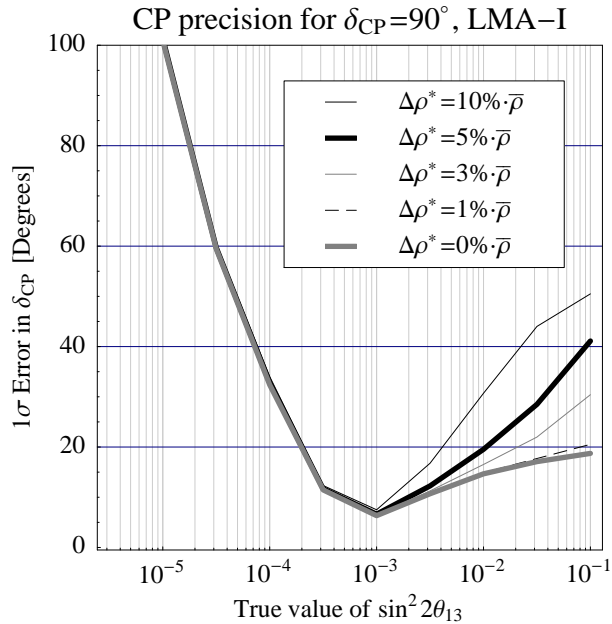}
\end{center}
\caption{\label{fig:mdu} CP precision as function of $\stheta$ for different values of the
matter density uncertainties. Figure from \Ref~\cite{Ohlsson:2003ip} for a $50 \, \mathrm{GeV}$
neutrino factory at $L=3000 \, \mathrm{km}$.}
\end{figure}

Expanded in small values of $\alpha \equiv \frac{\Delta m_{21}^2}{\Delta m_{31}^2} \simeq \pm 0.03$
and $\sin 2 \theta_{13}$ up to second order, the appearance probability $\nu_e \leftrightarrow \nu_\mu$ can be approximated by~\cite{Cervera:2000kp,Freund:2000ti,Freund:2001pn}
\begin{eqnarray}
P_{\mathrm{app}} & \simeq &  \sin^2 2 \theta_{13}  \, \sin^2 \theta_{23} \frac{\sin^2[(1-\hat{A}){\Delta}]}{(1-\hat{A})^2}
\nonumber \\
&\pm&  \alpha  \, \sin 2 \theta_{13} \,  \sin 2\theta_{12}  \sin 2\theta_{23} \, \sin \delta_{\mathrm{CP}}
\sin({\Delta})  \frac{\sin(\hat{A}{\Delta})}{\hat{A}}  \frac{\sin[(1-\hat{A}){\Delta}]}{(1-\hat{A})}
\nonumber  \\
&+&     \alpha  \, \sin 2 \theta_{13}  \,  \sin 2\theta_{12}  \sin 2\theta_{23} \, \cos \delta_{\mathrm{CP}} \cos({\Delta})  \frac{\sin(\hat{A}{\Delta})}{\hat{A}}  \frac{\sin[(1-\hat{A}){\Delta}]} {(1-\hat{A})}
 \nonumber  \\
&+&   \alpha^2   \, \cos^2 \theta_{23}  \sin^2 2\theta_{12} \frac{\sin^2(\hat{A}{\Delta})}{\hat{A}^2} \, ,
\label{equ:Papp}
\end{eqnarray}
where $\Delta \equiv \frac{\Delta m_{31}^2 L}{4 E}$ and $\hat{A} \equiv \pm \frac{2 \sqrt{2} G_F n_e E}{\Delta m_{31}^2}$. This probability is sensitive to the most interesting parameters: $\stheta$, $\deltacp$, and the mass hierarchy (via $\hat{A}$). However, because of this quite complicated structure, connected (``correlations'') and disconnected (``degeneracies'') degenerate solutions in the parameter space (at the chosen $\Delta \chi^2$) affect the extraction of the individual parameters. 
For the disconnected (discrete) degeneracies, we know the octant $(\theta_{23},\pi/2 - \theta_{23})$ degeneracy~\cite{Fogli:1996pv},
the $\mathrm{sgn}(\ldm)$-degeneracy~\cite{Minakata:2001qm}, and the intrinsic $(\theta_{13},\deltacp)$-degeneracy~\cite{Burguet-Castell:2001ez}, leading to an overall ``eight-fold'' degeneracy~\cite{Barger:2001yr}. 
\begin{figure}[t]
\begin{center}
\includegraphics[width=7.5cm,angle=-90]{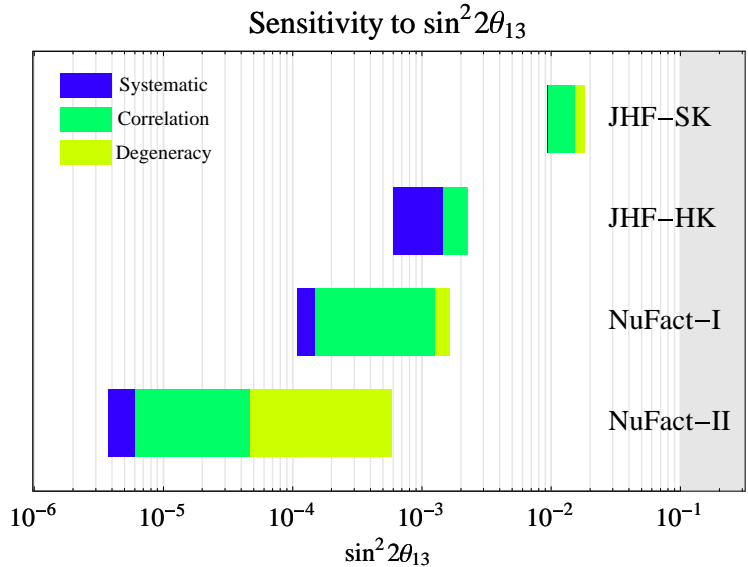}
\end{center}
\caption{\label{fig:exec2} Sensitivity to $\stheta$ at the 90\% confidence level for a superbeam (JHF-SK, corresponds to T2K), a superbeam upgrade (JHF-HK, corresponds to T2K with Hyper-Kamiokande detector and $4 \, \mathrm{MW}$ proton driver upgrade), a small neutrino factory (NuFact-I, integrated luminosity $\mathcal{L} = 10^{22}$ useful muon decays $\times$ $\mathrm{kt}$ detector mass), and a large neutrino factory (NuFact-II, $\mathcal{L} = 4 \cdot 10^{23}$ useful muon decays $\times$ $\mathrm{kt}$ detector mass).
Systematics, correlations, and degeneracies are successively included when moving from the left to right edges of the bars. Figure from \Ref~\cite{Huber:2002mx}.}
\end{figure}
In addition, it is well known that matter density uncertainties, \ie, correlations with the
matter density, challenge the extraction of the neutrino oscillation parameters (see, \eg, \Refs~\cite{Jacobsson:2001zk,Geller:2001ix,Shan:2001br,Fogli:2001tm,Ota:2002fu,Shan:2003vh,Kozlovskaya:2003kk,Ohlsson:2003ip} and references therein). As it can be read off from seismic wave reconstructions of the Earth's mantle density, uncertainties of the order of $5\%$ 
have to be assumed~\cite{Geller:2001ix}. One can easily see in \figu{mdu} that such uncertainties 
highly affect the measurement of $\deltacp$ (and, similarly, $\stheta$) especially for large values of $\stheta$. As it is obvious from this figure, a matter density precision of better than $1\%$ would eliminate the correlations with the matter density. Eventually, the full impact of all correlations and degenerate solutions was first demonstrated in \Refs~\cite{Freund:2001ui,Huber:2002mx}, and can by far exceed the impact of systematics. 
As one can read off \figu{exec2}, the impact of correlations and degeneracies becomes worse
for larger experiments, because these experiments are less dominated by (poor) statistics.
Therefore, strategies to resolve degenerate solutions will be needed for 
neutrino factories.

\subsection{How to quantify the neutrino factory performance?}

\begin{table}[t!]
\caption{\label{tab:cpperf} A number of performance indicators for $\delta_{\mathrm{CP}}$.
The level of condensation and the necessary computation power increase from the top to the
bottom of this table.}
\begin{center}
\begin{tabular}{||p{4.2cm}|p{10cm}||}
\hline
\hline
Performance indicator & Purpose \\
\hline
\hline
Allowed region in $\theta_{13}$-$\delta_{\mathrm{CP}}$-plane & Identify how much parameter
space remains for a specific assumption of simulated values \\
\hline
Sensitivity to max. CP violation & Show range in which max. CP violation ($\delta_{\mathrm{CP}} = \pi/2$ or $3 \pi/2$) can be detected (typically shown as function of $\stheta$ and third parameter) \\
\hline
Sensitivity to ``any'' CP violation & Show range in which ``any'' CP violation ($\delta_{\mathrm{CP}} \neq 0$ and $\pi$) can be detected (typically shown as function of $\stheta$ and $\deltacp$ or ``Fraction of $\deltacp$'') \\
\hline
Precision of $\delta_{\mathrm{CP}}$ & Show how precisely $\delta_{\mathrm{CP}}$ can be measured. Problem: Only defined in high-precision limit ($\delta_{\mathrm{CP}}$ cyclic, not Gaussian) \\
\hline
CP coverage & Show which fraction of possible values of $\delta_{\mathrm{CP}}$ fit a chosen simulated
value (small values mean high precision, whereas $360^\circ$ mean no information). Describes high precision and exclusion measurements for all possible values of $\delta_{\mathrm{CP}}$ (typically shown as function of $\stheta$ or $\deltacp$) \\
\hline
\hline
\end{tabular}
\end{center}
\end{table}

In order to quantify and optimize the neutrino factory performance, performance indicators are
needed. The choice of performance indicator often depends on the tested hypothesis, the
purpose, computation power, and is in fact a matter of taste, too. For example, in \Tab~\ref{tab:cpperf}, a number of performance indicators for $\delta_{CP}$ are listed. 
Naturally, the computation time increases with going down in this table.
Compared to $\stheta$, the measurement of $\deltacp$ is more difficult to visualize because of
an extra degree of freedom.\footnote{For example, the $\stheta$ sensitivity depends on the
simulated and fit values of $\deltacp$. However, because (depending on the definition) either
the simulated (for the exclusion limit) or fit (for the discovery potential) $\stheta$ is zero, the corresponding value of $\deltacp$ is meaningless. Therefore, the $\stheta$ limits are simpler by one degree of freedom than any $\deltacp$ indicator, because the latter always depends on both the simulated and fit values of
$\deltacp$.} Therefore, for risk minimization purposes, a high level of condensation is necessary (bottom of table), whereas for the visualization of figures as close as possible to the actual results, a lower level of condensation is more illustrative (top of table). A good compromise is usually
the sensitivity to maximal/any CP violation (middle) with a moderate computation effort.
However, these performance indicators do not demonstrate that coincident degeneracies in
specific regions of the parameter space can reduce the $\deltacp$ precision by a factor of five compared to the optimum, because (compared to superbeams) this happens off the most often discussed values $\delta_{\mathrm{CP}}=0$, $\pi/2$, $\pi$, and $3 \pi/2$. This can only seen with the ``CP coverage''~\cite{Winter:2003ye,Huber:2004gg} (see dark curve in \figu{patmagic}, close to
$\deltacp \sim 7/8 \pi$). 
 
\subsection{Strategies for degeneracy resolution}

\begin{figure}[t!]
\begin{center}
\includegraphics[width=8cm]{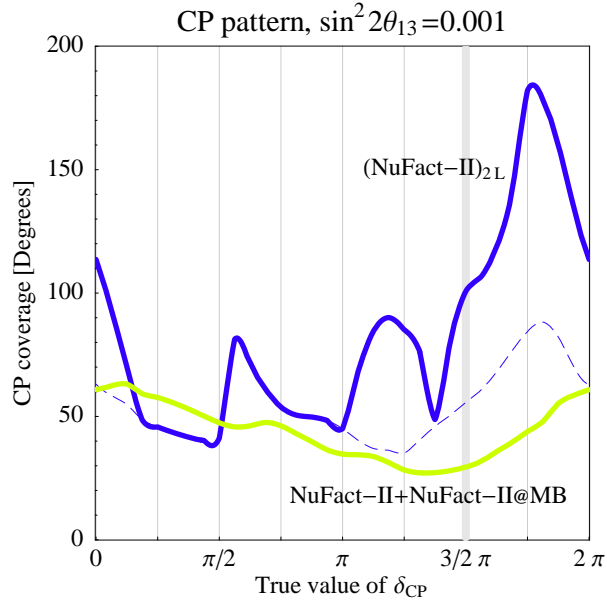}
\end{center}
\caption{\label{fig:patmagic} CP coverage for a neutrino factory with $E_\mu =50 \, \mathrm{GeV}$ as function of the simulated value of $\deltacp$ (``CP pattern'', $3 \sigma$). The label ``(NuFact-II)$_{(2L)}$'' refers to the double detector mass $100 \, \mathrm{kt}$ at $3 \, 000 \, \mathrm{km}$, whereas the label ``NuFact-II+NuFact-II@MB'' refers
to two baselines at $3 \, 000 \, \mathrm{km}$ and $7 \, 500 \, \mathrm{km}$ (``magic baseline'')
operated with a $50 \, \mathrm{kt}$ detector each. Dashed curves do not contain (disconnected) degeneracies. \fig\ from \Ref~\cite{Huber:2004gg}.}
\end{figure}

For the resolution of degenerate solutions at a neutrino factory, several methods have been
proposed in the literature. One possibility is the combination of a neutrino factory with a
superbeam upgrade~\cite{Burguet-Castell:2002qx}. An interesting approach for large values
of $\stheta$ may be the combination with the ``silver channel'' $\nu_e \rightarrow \nu_\tau$~\cite{Donini:2002rm,Autiero:2003fu}. In addition, better detectors with lower
thresholds may help to resolve the intrinsic degeneracy (see, \eg, \fig~27 of \Ref~\cite{Huber:2002mx}). Note that not all of these methods can be used in a wide range
of $\stheta$ values. A very powerful method, which has been demonstrated to work down
to $\stheta \sim 10^{-4}$, is the ``magic baseline''~\cite{Huber:2003ak}. As one can read off
\equ{Papp}, the condition $\sin(\hat{A} \Delta) \equiv 0$ forces all but the first term to
disappear, and therefore allows for a correlation- and degeneracy-free measurement of $\stheta$ and
the mass hierarchy. This condition evaluates to $\sqrt{2} G_F n_e L = 2 \pi$ independent of the
neutrino energy and the oscillation parameters, \ie, $L \sim 7 \, 000 - 7\, 500 \, \mathrm{km}$.
\figu{patmagic} illustrates how the magic baseline can be used for a risk-minimized measurement of $\deltacp$. One remaining issue
is the octant degeneracy provided that $\theta_{23}$ is substantially off maximal mixing:
A resolution of this degeneracy for very small values of $\stheta$ and $\theta_{23}$ very close to maximal mixing may be very difficult (see, \eg, \Ref~\cite{Huber:2005ep} for long-baseline data combined with atmospheric data).

\subsection{Physics case for a neutrino factory?}

Establishing the oscillation physics case for a neutrino factory is one of the major priorities of the currently ongoing ``International scoping study of a future neutrino factory and super-beam facility''~\cite{ISS}. For example, in terms of $\stheta$ and $\deltacp$, the physics cases
could look like this:
\begin{description}
\item[Large $\boldsymbol{\theta_{13}}$: $\boldsymbol{\sin^2 2 \theta_{13} \gtrsim 0.01}$] In this case,
the choice of technology may be the most relevant issue (superbeam versus beta beam versus neutrino factory). Current discussions show that it is not trivial to answer this question, because it depends on systematics, matter density uncertainties, and the neutrino factory detector performance. 
\item[Medium $\boldsymbol{\theta_{13}}$: $\boldsymbol{10^{-4} \lesssim \sin^2 2 \theta_{13} \lesssim 10^{-2}}$] This may be the ``golden age'' of the neutrino factory (or a higher gamma beta beam). Different sub-cases may, depending on $\stheta$, need special attention for the neutrino factory:
\begin{description}
\item[$\boldsymbol{0 \lesssim \delta_{\mathrm{CP}} \lesssim \pi}$] Few degeneracy problems. Use large luminosities at $L=3 \, 000 \, \mathrm{km}$?
\item[$\boldsymbol{\pi \lesssim \delta_{\mathrm{CP}} \lesssim 2 \pi}$] Many degeneracy problems. Use silver channels or ``magic baseline''?
\end{description}
\item[``Zero'' $\boldsymbol{\theta_{13}}$: $\boldsymbol{\sin^2 2 \theta_{13} \ll 10^{-4}}$] This case implies that $\stheta$ is well below the reach of a neutrino factory. Important questions are: What does this mean theoretically? What other physics applications can still be done in this case?
\end{description}
For example, for the physics case $\theta_{13}=0$, interesting applications may be establishing the MSW effect in Earth matter through the solar appearance term (fourth term in \equ{Papp})~\cite{Winter:2004mt}, or determining the mass hierarchy through a high statistics disappearance measurement~\cite{deGouvea:2005hk,deGouvea:2005mi}. In all of these cases, a ``very long'' neutrino factory baseline with $L \gg 3 \, 000 \, \mathrm{km}$ is required.

\subsection{Optimization of the physics potential}

Except from technical challenges, such as the detection system, the optimization of the
physics potential of a neutrino factory in terms of baseline(s), muon energies, and
required oscillation channels is an important subject to determine the layout for an
optimal neutrino factory. Earlier studies for the
optimization in $L$-$E_{\mu}$-space include, for instance, \Refs~\cite{Barger:1999jj,Freund:2001ui,Bueno:2001jd}. However, these studies did not take into
account the full degeneracy problem because it was either not fully known yet, or technically not fully accessible. More recent efforts will be devoted to this problem~\cite{ISS,ISSProject}. For example, one can show that after the inclusion
of degeneracies, the ``magic baseline'' gives the best $\stheta$ sensitivity as opposed to shorter baselines, which are optimal for the systematics limit only~\cite{ISSProject}. In addition, somewhat lower muon
energies than $50 \, \mathrm{GeV}$ might be acceptable, which could be further reduced by
an improved detection system. However, the sometimes discussed option of an initial $L \sim 1 \, 000 \, \mathrm{km}$/$E_{\mu} = 20 \, \mathrm{GeV}$ neutrino factory will be far off the optimum. As far as the channel requirements are concerned, the ``golden'' channels $\nu_e \rightarrow \nu_\mu$ and $\bar{\nu}_e \rightarrow \bar{\nu}_\mu$ will be the most interesting ones. However, it has been emphasized that the disappearance information $\nu_\mu \rightarrow \nu_\mu$ (or $\bar{\nu}_\mu \rightarrow \bar{\nu}_\mu$) is important to improve the precisions on the leading atmospheric parameters since these translate into uncertainties in $\stheta$ and $\deltacp$ via multi-parameter correlations~\cite{Huber:2002mx,Donini:2005db,Donini:2005gy}. As demonstrated in \Refs~\cite{deGouvea:2005mi,ISSProject}, using a different data set without charge identification 
for the disappearance channel improves the results tremendously. In this case, higher efficiencies are obtained for the price of adding the ``wrong-sign'' and ``right-sign'' muon events.
Eventually, the requirement and optimization of ``silver'' ($\nu_e \rightarrow \nu_\tau$/$\bar{\nu}_e \rightarrow \bar{\nu}_\tau$)
and ``platinum'' ($\nu_\mu \rightarrow \nu_e$/$\bar{\nu}_\mu \rightarrow \bar{\nu}_e$) channels needs further investigation~\cite{ISSProject}.

\subsection{Other oscillation physics: New physics tests and neutrino geophysics}

Beyond neutrino oscillations, there could be ad-mixtures of 
other ``new physics'' in addition to neutrino oscillations. Such effects can be motivated by
neutrino decay, decoherence, sterile neutrinos, lepton flavor violation, extra
dimensions, mass-varying neutrinos, 
or others. For a neutrino factory, it will therefore be important to
test the consistency of the oscillation hypothesis at the precision level.
 Except from antineutrino running,
which is usually included in neutrino factory studies, there are a number of conceptual
approaches to this problem:
\begin{description}
\item[New channels,] such as using $\nu_\tau$ detection (``silver channel''~\cite{Donini:2002rm,Autiero:2003fu}), may be a key component
to test unitarity relationships as function
of energy, such as $P_{ee} + P_{e \mu} + P_{e \tau} = 1$.
\item[Neutral current measurements] are hard because of cross section uncertainties
and backgrounds~\cite{Barger:2004db}. However, they still might provide valuable
information if the cross section were better known and the CC/NC event selection was
improved. Note that neutral currents can only access a subset of new physics, such as 
sterile neutrinos or (invisible) neutrino decay.
\item[Unitarity triangles] similar to the quark sector might be used~\cite{Farzan:2002ct,Zhang:2004hf,Xing:2005gk}. However, extracting the
relevant angles and sides will be a challenging task involving many experiments~\cite{Farzan:2002ct}.
\item[Spectral signatures,] which are characteristic for specific effects, might be tested~\cite{Blennow:2005yk}.
In particular, effects on the probability level (decoherence, decay, \etc) lead to a depletion or
enhancement in specific regions of the spectrum, while the oscillation nodes remain more or less
unchanged.
\item[Hamiltonian-level effects,]  such as from lepton flavor violating non-standard interactions
or mass-varying neutrinos, may be hardest to access because they shift the oscillation pattern
(see, \eg, \Ref~\cite{Blennow:2005qj}). As a consequence, the confusion with the standard
oscillation parameters is a major issue (see, \eg, \Ref~\cite{Huber:2002bi}).
\end{description}
It is an often suggested strategy just to assume ``standard'' three-flavor neutrino oscillations
until an inconsistency is discovered. Note, however, that some of the approaches
above require action beforehand, such as the silver channel measurement.

Eventually, a very different direction for a neutrino factory might be geophysics applications.
For instance, it has been demonstrated in \Ref~\cite{Winter:2005we} that using the MSW
effect, a neutrino factory could measure the Earth's inner core density at the per cent level for $\stheta \gtrsim 0.01$, where the correlations with the unknown oscillation parameters were taken into account.\footnote{The CP and solar terms in \equ{Papp} are suppressed
compared to the first term for a very long neutrino factory baseline when the $1/L^2$-dependence of the flux is taken into account (see, \eg, \Ref~\cite{Winter:2005ud}). Therefore, the uncertainties in the oscillation parameters (such as $\deltacp$)
have relatively little effect.} For a recent review on neutrino tomography, see \Ref~\cite{Winter:2006vg} and references therein.

\section{Beta beam}

Beta beams~\cite{Zucchelli:2002sa} were originally proposed for the CERN layout~\cite{Autin:2002ms,Mezzetto:2003ub,Bouchez:2003fy,Lindroos:2003kp,Lindroos:2004sa}.
A beta beam uses the beta decays of ions in straight sections of a storage ring
to produce a neutrino beam. Since the
beta decays only produce the electron flavor, there is, compared to a neutrino factory, no need for charge identification in the detector. In addition, compared to superbeams, there is no intrinsic beam background
limiting the beta beams, which effectively means that there is no limitation to $\stheta \gtrsim 0.001$. Since one could, in principle, accelerate the ions to fairly high energies (gammas)~\cite{Burguet-Castell:2003vv},
beta beams could, depending on the gamma, be interesting alternatives to either superbeams or
neutrino factories. A major technical challenge is
to produce sufficiently large numbers of ions, typically $^6_2$He (for antineutrinos)
and $^{18}_{10}$Ne (for neutrinos). These ions have to be accelerated and then injected into the storage ring, where they decay. For the SPL at CERN, often used gammas are $150$ (for $^6_2$He) and $60$ (for $^{18}_{10}$Ne). For the number of useful ion decays per year, usually of the order of $3 \cdot 10^{18}$ (for $^6_2$He) and $1 \cdot 10^{18}$ (for $^{18}_{10}$Ne) are assumed.
Most of recent beta beams studies are oriented towards these numbers of useful ion decays per year.

\subsection{Beta beam as function of gamma}

As mentioned above, beta beams can have different purposes depending on gamma factor and
optimization. The following is an attempt to classify possible beta beam scenarios 
according to required equipment and purpose:
\begin{description}
\item[``Very low'' gamma ($\boldsymbol{\gamma < 150?}$)]~\cite{Zucchelli:2002sa,Autin:2002ms,Mezzetto:2003ub,Bouchez:2003fy,Lindroos:2003kp,ISS,Couce} This range was originally proposed for the CERN layout (SPS) and could be an alternative to superbeam upgrades. Usually, a Water Cherenkov detector is proposed. In addition to neutrino oscillations, neutrino-nucleon interactions and a possible neutrino magnetic moment may be tested for very small gamma~\cite{Volpe:2003fi,McLaughlin:2003yg,Serreau:2004kx}. 
\item[``Low'' gamma ($\boldsymbol{150 < \gamma < 350?}$)]~\cite{Burguet-Castell:2003vv,Burguet-Castell:2005pa,Huber:2005jk,ISS,Couce} Such a beta beam might be possible at an upgraded SPS. The physics potential is, to current knowledge, competitive to superbeam upgrades. Typically a Water Cherenkov detector is used.
\item[``Medium'' gamma ($\boldsymbol{350 < \gamma < 800?}$)]~\cite{Burguet-Castell:2003vv,Huber:2005jk} For such setups, a rather large accelerator will be needed (Tevatron-size?). The detector technology might be possibly Water Cherenkov or TASD, or another. The physics potential probably lies between superbeam upgrade and
neutrino factory.
\item[``High'' gamma ($\boldsymbol{\gamma > 800?}$)]~\cite{Burguet-Castell:2003vv,Huber:2005jk,Agarwalla:2005we} This could be an alternative to a neutrino factory in terms of physics potential. However, a very large accelerator (LHC-size?) will be needed. For the detector, a technology different than Water is required due to the domination of non-QE events.
\end{description}
In summary, the gamma determines the neutrino energies and therefore the required detection
technology: Both physics potential and effort increase drastically with increasing gamma.

\subsection{Optimization of a beta beam}

\begin{table}[t!]
\caption{\label{tab:setups} Some experiment representatives used for the comparison
of different experiments. Table similar to \Ref~\cite{Huber:2005jk}.}
\begin{center}
\begin{tabular}{||l|r|r|r|l|r|r|c||}
\hline
\hline
Label & $\gamma$ & \Large{$\frac{L}{\mathrm{km}}$} & \Large{$\frac{\langle E_{\nu} \rangle}{\mathrm{GeV}}$} & Detector & \Large{$\frac{m_{\mathrm{Det}}}{\mathrm{kt}}$} & {\Large $\frac{t_{\mathrm{run}}}{\mathrm{yr}}$}  $(\nu,\bar{\nu})$ & \Ref \\
\hline
Setup~1 & 200 & 520 & 0.75 & Water Cherenkov & 500 & (4,4) & \cite{Huber:2005jk} \\
Setup~2 & 500 & 650 & 1.9 & TASD & 50 & (4,4) & \cite{Huber:2005jk} \\
Setup~3 & 1000 & 1300 & 3.8 & TASD & 50 & (4,4) & \cite{Huber:2005jk}  \\
\hline
\JHFHK\ & n/a & 295 & 0.76 & Water Cherenkov & 500 & (2,6) & \cite{Huber:2002mx} \\
NF@3000km & n/a & 3000 & 33 & Magn. iron calor. & 50 & (4,4) & \cite{Huber:2002mx} \\
NF@7500km & n/a & 7500 & 33 & Magn. iron calor. & 50 & (4,4) & \cite{Huber:2003ak} \\
\hline
\hline
\end{tabular}
\end{center}
\end{table}

\begin{figure}[t!]
\begin{center}
\includegraphics[width=\textwidth]{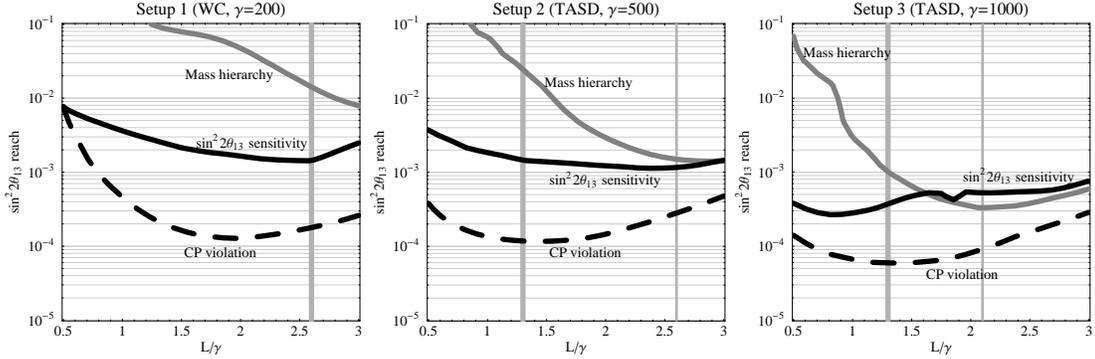}
\end{center}
\caption{\label{fig:lcomp} The optimization of the sensitivity to the normal mass hierarchy, the $\stheta$ sensitivity, and the sensitivity to maximal CP violation $\deltacp=\pi/2$ as function of baseline (fixed gamma in each panel, $3 \sigma$). The setups are defined in \Tab~\ref{tab:setups}. Figure taken from \Ref~\cite{Huber:2005jk}.}
\end{figure}

The optimization of beta beams includes many factors, such as baseline, gamma factor, neutrino and antineutrino luminosities (and the relative gamma), and detector technology (see, \eg, \Refs~\cite{Burguet-Castell:2003vv,Burguet-Castell:2005pa,Huber:2005jk}). As far as the
overall gamma is concerned, it determines the accelerator size, detection technology, and
purpose of the experiment, as we have discussed in the last section. In general, for the physics potential, the rule ``the larger the gamma, the better'' applies -- provided that the detector technology is chosen accordingly~\cite{Burguet-Castell:2003vv,Huber:2005jk}. For the relative gamma between neutrinos and antineutrinos, there is no obvious gain by increasing one of the two gammas~\cite{Burguet-Castell:2005pa}. One can also change the relative neutrino-antineutrino running time (or luminosity) without severe loss in sensitivities as long as it is not reduced under about 20\% of the total running time~\cite{Huber:2005jk}. Since the choice of
gamma is rather one of the technology and purpose, the optimization of the baseline for a specific (fixed) gamma is probably the most interesting physics optimization question for a beta beam. In \figu{lcomp}, several setups with fixed gammas and specific detectors are shown, where the sensitivities to mass hierarchy, $\stheta$, and CP violation are optimized (including correlations and degeneracies; for further degeneracy studies, see \Refs~\cite{Donini:2004hu,Donini:2004iv}). It is not very surprising that long baselines help for the mass hierarchy, whereas shorter baselines are preferable for CP violation. However, it is interesting that, for low gamma, long baselines help for the $\stheta$ sensitivity, because correlations and degeneracies are reduced in the second oscillation maximum. In summary, the optimal baseline depends on what quantity one is optimizing for. Similar to the neutrino factory, two baselines may be finally optimal.

\subsection{Comparison to superbeam and neutrino factory}

\begin{figure}[t!]
\begin{center}
\includegraphics[width=9cm]{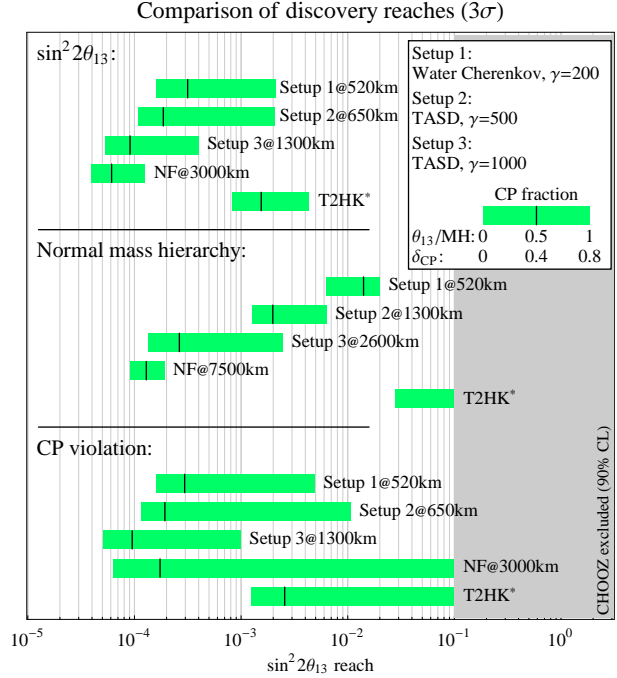}
\end{center}
\caption{\label{fig:ds} Discovery reach comparison at the $3 \sigma$ confidence level,
where the bars represent the optimal $\deltacp$ (left edge), the ``typical'' $\deltacp$ (vertical line), and the most conservative $\deltacp$ (right edge) realized by nature. The setups are defined in \Tab~\ref{tab:setups} and the baselines are chosen to compare optimized setups 
(especially for the mass hierarchy). Figure taken from \Ref~\cite{Huber:2005jk}.}
\end{figure}

For a beta beam, the comparison to a superbeam upgrade and a neutrino
factory may be very important to evaluate the physics potential. In \figu{ds},
a comparison of the $\stheta$, mass hierarchy, and CP violation sensitivity discovery reaches
is shown, where the setups from \Tab~\ref{tab:setups} are used. The bars represent the
possible $\stheta$ range depending on the value of $\deltacp$ chosen by nature.
There are several interesting observations: First, even the lower gamma beta beam option 
can easily compete with the superbeam upgrade. Second, the neutrino factory, if optimized
for the chosen quantity, is the best option for the $\stheta$ and mass hierarchy sensitivities,
as well as it is comparable to the beta beams for the ``typical'' $\deltacp$. However, all of the beta beams have excellent sensitivities to CP violation. These results, of course, depend on the achievable luminosities and systematics. In addition, note that the degeneracy problem at the
neutrino factory can be reduced by the combination of the two baselines. In conclusion,
beta beams could, for large enough gamma, at least theoretically compete with neutrino factories. Further studies will demonstrate which concept is more feasible (see, \eg, \Ref~\cite{ISS}).

\section{Outlook}

The current key questions for a neutrino factory complex are discussed
in the ongoing ISS study~\cite{ISS}. For example, the choice of the experimental
program for large values of $\stheta$ (superbeam, beta beam, or neutrino factory),
and the requirements to a neutrino factory
in that case are very important issues which depend on many variables, such as
matter density uncertainties, the detector optimization, and systematics issues. 
In addition, the feasibility of technical
aspects (muon cooling, target power, storage ring layout, \etc) as well as the
optimization of the detection system are under investigation. For a beta beam,
the number of achievable ion decays seems to be the most critical issue.
Answers may be expected from the EURISOL design study~\cite{EURISOL}. In addition,
as far as the selection between neutrino factory and beta beam is concerned, 
the required effort for the beta beam in terms of storage ring and accelerator size 
 in order to achieve a similar physics potential needs further study. On the theoretical
side, the justification of a future high precision neutrino oscillation program
requires a solid justification in terms of the physics cases and the meaning for
theory. The ongoing studies will certainly provide answers to many of these questions,
and direction for future research. 

\section{Acknowledgments}
I would to thank the organizers for inviting me to NO-VE 2006,
as well as I would like to congratulate to an excellent organization.
In addition, I want to
acknowledge support from the W.~M.~Keck Foundation and NSF grant PHY-0503584.


\begin{thebibliography}{10}
\expandafter\ifx\csname bibnamefont\endcsname\relax
  \def\bibnamefont#1{#1}\fi
\expandafter\ifx\csname bibfnamefont\endcsname\relax
  \def\bibfnamefont#1{#1}\fi
\expandafter\ifx\csname url\endcsname\relax
  \def\url#1{\texttt{#1}}\fi
\expandafter\ifx\csname urlprefix\endcsname\relax\def\urlprefix{URL }\fi
\providecommand{\bibinfo}[2]{#2}
\providecommand{\eprint}[2][]{\url{#2}}

\bibitem{Geer:1998iz}
\bibinfo{author}{\bibfnamefont{S.}~\bibnamefont{Geer}}, \bibinfo{journal}{Phys.
  Rev.} \textbf{\bibinfo{volume}{D57}}, \bibinfo{pages}{6989}
  (\bibinfo{year}{1998}), \eprint[http://arXiv.org/abs]{hep-ph/9712290}.

\bibitem{Zucchelli:2002sa}
\bibinfo{author}{\bibfnamefont{P.}~\bibnamefont{Zucchelli}},
  \bibinfo{journal}{Phys. Lett.} \textbf{\bibinfo{volume}{B532}},
  \bibinfo{pages}{166} (\bibinfo{year}{2002}).

\bibitem{Burguet-Castell:2003vv}
\bibinfo{author}{\bibfnamefont{J.}~\bibnamefont{Burguet-Castell}},
  \bibinfo{author}{\bibfnamefont{D.}~\bibnamefont{Casper}},
  \bibinfo{author}{\bibfnamefont{J.~J.} \bibnamefont{Gomez-Cadenas}},
  \bibinfo{author}{\bibfnamefont{P.}~\bibnamefont{Hernandez}},
  \bibnamefont{and} \bibinfo{author}{\bibfnamefont{F.}~\bibnamefont{Sanchez}},
  \bibinfo{journal}{Nucl. Phys.} \textbf{\bibinfo{volume}{B695}},
  \bibinfo{pages}{217} (\bibinfo{year}{2004}), \eprint{hep-ph/0312068}.

\bibitem{Albright:2000xi}
\bibinfo{author}{\bibfnamefont{C.}~\bibnamefont{Albright}} \emph{et~al.}
  \eprint{hep-ex/0008064}.

\bibitem{Apollonio:2002en}
\bibinfo{author}{\bibfnamefont{M.}~\bibnamefont{Apollonio}} \emph{et~al.}
  \eprint[http://arXiv.org/abs]{hep-ph/0210192}.

\bibitem{DeRujula:1998hd}
\bibinfo{author}{\bibfnamefont{A.}~\bibnamefont{De~Rujula}},
  \bibinfo{author}{\bibfnamefont{M.~B.} \bibnamefont{Gavela}},
  \bibnamefont{and}
  \bibinfo{author}{\bibfnamefont{P.}~\bibnamefont{Hernandez}},
  \bibinfo{journal}{Nucl. Phys.} \textbf{\bibinfo{volume}{B547}},
  \bibinfo{pages}{21} (\bibinfo{year}{1999}), \eprint{hep-ph/9811390}.

\bibitem{Barger:1999jj}
\bibinfo{author}{\bibfnamefont{V.~D.} \bibnamefont{Barger}},
  \bibinfo{author}{\bibfnamefont{S.}~\bibnamefont{Geer}},
  \bibinfo{author}{\bibfnamefont{R.}~\bibnamefont{Raja}}, \bibnamefont{and}
  \bibinfo{author}{\bibfnamefont{K.}~\bibnamefont{Whisnant}},
  \bibinfo{journal}{Phys. Rev.} \textbf{\bibinfo{volume}{D62}},
  \bibinfo{pages}{013004} (\bibinfo{year}{2000}), \eprint{hep-ph/9911524}.

\bibitem{Cervera:2000kp}
\bibinfo{author}{\bibfnamefont{A.}~\bibnamefont{Cervera}} \emph{et~al.},
  \bibinfo{journal}{Nucl. Phys.} \textbf{\bibinfo{volume}{B579}},
  \bibinfo{pages}{17} (\bibinfo{year}{2000}), \eprint{hep-ph/0002108}.

\bibitem{ISS}
\emph{\bibinfo{title}{International scoping study of a future neutrino factory
  and super-beam facility}}, \bibinfo{note}{{\tt
  http://www.hep.ph.ic.ac.uk/iss/}}.

\bibitem{Ohlsson:2003ip}
\bibinfo{author}{\bibfnamefont{T.}~\bibnamefont{Ohlsson}} \bibnamefont{and}
  \bibinfo{author}{\bibfnamefont{W.}~\bibnamefont{Winter}},
  \bibinfo{journal}{Phys. Rev.} \textbf{\bibinfo{volume}{D68}},
  \bibinfo{pages}{073007} (\bibinfo{year}{2003}), \eprint{hep-ph/0307178}.

\bibitem{Freund:2000ti}
\bibinfo{author}{\bibfnamefont{M.}~\bibnamefont{Freund}},
  \bibinfo{author}{\bibfnamefont{P.}~\bibnamefont{Huber}}, \bibnamefont{and}
  \bibinfo{author}{\bibfnamefont{M.}~\bibnamefont{Lindner}},
  \bibinfo{journal}{Nucl. Phys.} \textbf{\bibinfo{volume}{B585}},
  \bibinfo{pages}{105} (\bibinfo{year}{2000}), \eprint{hep-ph/0004085}.

\bibitem{Freund:2001pn}
\bibinfo{author}{\bibfnamefont{M.}~\bibnamefont{Freund}},
  \bibinfo{journal}{Phys. Rev.} \textbf{\bibinfo{volume}{D64}},
  \bibinfo{pages}{053003} (\bibinfo{year}{2001}), \eprint{hep-ph/0103300}.

\bibitem{Fogli:1996pv}
\bibinfo{author}{\bibfnamefont{G.~L.} \bibnamefont{Fogli}} \bibnamefont{and}
  \bibinfo{author}{\bibfnamefont{E.}~\bibnamefont{Lisi}},
  \bibinfo{journal}{Phys. Rev.} \textbf{\bibinfo{volume}{D54}},
  \bibinfo{pages}{3667} (\bibinfo{year}{1996}), \eprint{hep-ph/9604415}.

\bibitem{Minakata:2001qm}
\bibinfo{author}{\bibfnamefont{H.}~\bibnamefont{Minakata}} \bibnamefont{and}
  \bibinfo{author}{\bibfnamefont{H.}~\bibnamefont{Nunokawa}},
  \bibinfo{journal}{JHEP} \textbf{\bibinfo{volume}{10}}, \bibinfo{pages}{001}
  (\bibinfo{year}{2001}), \eprint[http://arXiv.org/abs]{hep-ph/0108085}.

\bibitem{Burguet-Castell:2001ez}
\bibinfo{author}{\bibfnamefont{J.}~\bibnamefont{Burguet-Castell}},
  \bibinfo{author}{\bibfnamefont{M.~B.} \bibnamefont{Gavela}},
  \bibinfo{author}{\bibfnamefont{J.~J.} \bibnamefont{Gomez-Cadenas}},
  \bibinfo{author}{\bibfnamefont{P.}~\bibnamefont{Hernandez}},
  \bibnamefont{and} \bibinfo{author}{\bibfnamefont{O.}~\bibnamefont{Mena}},
  \bibinfo{journal}{Nucl. Phys.} \textbf{\bibinfo{volume}{B608}},
  \bibinfo{pages}{301} (\bibinfo{year}{2001}),
  \eprint[http://arXiv.org/abs]{hep-ph/0103258}.

\bibitem{Barger:2001yr}
\bibinfo{author}{\bibfnamefont{V.}~\bibnamefont{Barger}},
  \bibinfo{author}{\bibfnamefont{D.}~\bibnamefont{Marfatia}}, \bibnamefont{and}
  \bibinfo{author}{\bibfnamefont{K.}~\bibnamefont{Whisnant}},
  \bibinfo{journal}{Phys. Rev.} \textbf{\bibinfo{volume}{D65}},
  \bibinfo{pages}{073023} (\bibinfo{year}{2002}),
  \eprint[http://arXiv.org/abs]{hep-ph/0112119}.

\bibitem{Huber:2002mx}
\bibinfo{author}{\bibfnamefont{P.}~\bibnamefont{Huber}},
  \bibinfo{author}{\bibfnamefont{M.}~\bibnamefont{Lindner}}, \bibnamefont{and}
  \bibinfo{author}{\bibfnamefont{W.}~\bibnamefont{Winter}},
  \bibinfo{journal}{Nucl. Phys.} \textbf{\bibinfo{volume}{B645}},
  \bibinfo{pages}{3} (\bibinfo{year}{2002}), \eprint{hep-ph/0204352}.

\bibitem{Jacobsson:2001zk}
\bibinfo{author}{\bibfnamefont{B.}~\bibnamefont{Jacobsson}},
  \bibinfo{author}{\bibfnamefont{T.}~\bibnamefont{Ohlsson}},
  \bibinfo{author}{\bibfnamefont{H.}~\bibnamefont{Snellman}}, \bibnamefont{and}
  \bibinfo{author}{\bibfnamefont{W.}~\bibnamefont{Winter}},
  \bibinfo{journal}{Phys. Lett.} \textbf{\bibinfo{volume}{B532}},
  \bibinfo{pages}{259} (\bibinfo{year}{2002}), \eprint{hep-ph/0112138}.

\bibitem{Geller:2001ix}
\bibinfo{author}{\bibfnamefont{R.~J.} \bibnamefont{Geller}} \bibnamefont{and}
  \bibinfo{author}{\bibfnamefont{T.}~\bibnamefont{Hara}},
  \bibinfo{journal}{Nucl. Instrum. Meth.} \textbf{\bibinfo{volume}{A503}},
  \bibinfo{pages}{187} (\bibinfo{year}{2001}), \eprint{hep-ph/0111342}.

\bibitem{Shan:2001br}
\bibinfo{author}{\bibfnamefont{L.-Y.} \bibnamefont{Shan}},
  \bibinfo{author}{\bibfnamefont{B.-L.} \bibnamefont{Young}}, \bibnamefont{and}
  \bibinfo{author}{\bibfnamefont{X.-m.} \bibnamefont{Zhang}},
  \bibinfo{journal}{Phys. Rev.} \textbf{\bibinfo{volume}{D66}},
  \bibinfo{pages}{053012} (\bibinfo{year}{2002}),
  \eprint[http://arXiv.org/abs]{hep-ph/0110414}.

\bibitem{Fogli:2001tm}
\bibinfo{author}{\bibfnamefont{G.~L.} \bibnamefont{Fogli}},
  \bibinfo{author}{\bibfnamefont{G.}~\bibnamefont{Lettera}}, \bibnamefont{and}
  \bibinfo{author}{\bibfnamefont{E.}~\bibnamefont{Lisi}}
  (\bibinfo{year}{2001}), \eprint{hep-ph/0112241}.

\bibitem{Ota:2002fu}
\bibinfo{author}{\bibfnamefont{T.}~\bibnamefont{Ota}} \bibnamefont{and}
  \bibinfo{author}{\bibfnamefont{J.}~\bibnamefont{Sato}},
  \bibinfo{journal}{Phys. Rev.} \textbf{\bibinfo{volume}{D67}},
  \bibinfo{pages}{053003} (\bibinfo{year}{2003}), \eprint{hep-ph/0211095}.

\bibitem{Shan:2003vh}
\bibinfo{author}{\bibfnamefont{L.-Y.} \bibnamefont{Shan}} \emph{et~al.},
  \bibinfo{journal}{Phys. Rev.} \textbf{\bibinfo{volume}{D68}},
  \bibinfo{pages}{013002} (\bibinfo{year}{2003}), \eprint{hep-ph/0303112}.

\bibitem{Kozlovskaya:2003kk}
\bibinfo{author}{\bibfnamefont{E.}~\bibnamefont{Kozlovskaya}},
  \bibinfo{author}{\bibfnamefont{J.}~\bibnamefont{Peltoniemi}},
  \bibnamefont{and} \bibinfo{author}{\bibfnamefont{J.}~\bibnamefont{Sarkamo}}
  (\bibinfo{year}{2003}), \eprint{hep-ph/0305042}.

\bibitem{Freund:2001ui}
\bibinfo{author}{\bibfnamefont{M.}~\bibnamefont{Freund}},
  \bibinfo{author}{\bibfnamefont{P.}~\bibnamefont{Huber}}, \bibnamefont{and}
  \bibinfo{author}{\bibfnamefont{M.}~\bibnamefont{Lindner}},
  \bibinfo{journal}{Nucl. Phys.} \textbf{\bibinfo{volume}{B615}},
  \bibinfo{pages}{331} (\bibinfo{year}{2001}),
  \eprint[http://arXiv.org/abs]{hep-ph/0105071}.

\bibitem{Winter:2003ye}
\bibinfo{author}{\bibfnamefont{W.}~\bibnamefont{Winter}},
  \bibinfo{journal}{Phys. Rev.} \textbf{\bibinfo{volume}{D70}},
  \bibinfo{pages}{033006} (\bibinfo{year}{2004}), \eprint{hep-ph/0310307}.

\bibitem{Huber:2004gg}
\bibinfo{author}{\bibfnamefont{P.}~\bibnamefont{Huber}},
  \bibinfo{author}{\bibfnamefont{M.}~\bibnamefont{Lindner}}, \bibnamefont{and}
  \bibinfo{author}{\bibfnamefont{W.}~\bibnamefont{Winter}},
  \bibinfo{journal}{JHEP} \textbf{\bibinfo{volume}{05}}, \bibinfo{pages}{020}
  (\bibinfo{year}{2005}), \eprint{hep-ph/0412199}.

\bibitem{Burguet-Castell:2002qx}
\bibinfo{author}{\bibfnamefont{J.}~\bibnamefont{Burguet-Castell}},
  \bibinfo{author}{\bibfnamefont{M.~B.} \bibnamefont{Gavela}},
  \bibinfo{author}{\bibfnamefont{J.~J.} \bibnamefont{Gomez-Cadenas}},
  \bibinfo{author}{\bibfnamefont{P.}~\bibnamefont{Hernandez}},
  \bibnamefont{and} \bibinfo{author}{\bibfnamefont{O.}~\bibnamefont{Mena}},
  \bibinfo{journal}{Nucl. Phys.} \textbf{\bibinfo{volume}{B646}},
  \bibinfo{pages}{301} (\bibinfo{year}{2002}),
  \eprint[http://arXiv.org/abs]{hep-ph/0207080}.

\bibitem{Donini:2002rm}
\bibinfo{author}{\bibfnamefont{A.}~\bibnamefont{Donini}},
  \bibinfo{author}{\bibfnamefont{D.}~\bibnamefont{Meloni}}, \bibnamefont{and}
  \bibinfo{author}{\bibfnamefont{P.}~\bibnamefont{Migliozzi}},
  \bibinfo{journal}{Nucl. Phys.} \textbf{\bibinfo{volume}{B646}},
  \bibinfo{pages}{321} (\bibinfo{year}{2002}),
  \eprint[http://arXiv.org/abs]{hep-ph/0206034}.

\bibitem{Autiero:2003fu}
\bibinfo{author}{\bibfnamefont{D.}~\bibnamefont{Autiero}} \emph{et~al.},
  \bibinfo{journal}{Eur. Phys. J.} \textbf{\bibinfo{volume}{C33}},
  \bibinfo{pages}{243} (\bibinfo{year}{2004}), \eprint{hep-ph/0305185}.

\bibitem{Huber:2003ak}
\bibinfo{author}{\bibfnamefont{P.}~\bibnamefont{Huber}} \bibnamefont{and}
  \bibinfo{author}{\bibfnamefont{W.}~\bibnamefont{Winter}},
  \bibinfo{journal}{Phys. Rev.} \textbf{\bibinfo{volume}{D68}},
  \bibinfo{pages}{037301} (\bibinfo{year}{2003}), \eprint{hep-ph/0301257}.

\bibitem{Huber:2005ep}
\bibinfo{author}{\bibfnamefont{P.}~\bibnamefont{Huber}},
  \bibinfo{author}{\bibfnamefont{M.}~\bibnamefont{Maltoni}}, \bibnamefont{and}
  \bibinfo{author}{\bibfnamefont{T.}~\bibnamefont{Schwetz}},
  \bibinfo{journal}{Phys. Rev.} \textbf{\bibinfo{volume}{D71}},
  \bibinfo{pages}{053006} (\bibinfo{year}{2005}), \eprint{hep-ph/0501037}.

\bibitem{Winter:2004mt}
\bibinfo{author}{\bibfnamefont{W.}~\bibnamefont{Winter}},
  \bibinfo{journal}{Phys. Lett.} \textbf{\bibinfo{volume}{B613}},
  \bibinfo{pages}{67} (\bibinfo{year}{2005}), \eprint{hep-ph/0411309}.

\bibitem{deGouvea:2005hk}
\bibinfo{author}{\bibfnamefont{A.}~\bibnamefont{de~Gouvea}},
  \bibinfo{author}{\bibfnamefont{J.}~\bibnamefont{Jenkins}}, \bibnamefont{and}
  \bibinfo{author}{\bibfnamefont{B.}~\bibnamefont{Kayser}},
  \bibinfo{journal}{Phys. Rev.} \textbf{\bibinfo{volume}{D71}},
  \bibinfo{pages}{113009} (\bibinfo{year}{2005}), \eprint{hep-ph/0503079}.

\bibitem{deGouvea:2005mi}
\bibinfo{author}{\bibfnamefont{A.}~\bibnamefont{de~Gouvea}} \bibnamefont{and}
  \bibinfo{author}{\bibfnamefont{W.}~\bibnamefont{Winter}},
  \bibinfo{journal}{Phys. Rev.} \textbf{\bibinfo{volume}{D73}},
  \bibinfo{pages}{033003} (\bibinfo{year}{2006}), \eprint{hep-ph/0509359}.

\bibitem{Bueno:2001jd}
\bibinfo{author}{\bibfnamefont{A.}~\bibnamefont{Bueno}},
  \bibinfo{author}{\bibfnamefont{M.}~\bibnamefont{Campanelli}},
  \bibinfo{author}{\bibfnamefont{S.}~\bibnamefont{Navas-Concha}},
  \bibnamefont{and} \bibinfo{author}{\bibfnamefont{A.}~\bibnamefont{Rubbia}},
  \bibinfo{journal}{Nucl. Phys.} \textbf{\bibinfo{volume}{B631}},
  \bibinfo{pages}{239} (\bibinfo{year}{2002}),
  \eprint[http://arXiv.org/abs]{hep-ph/0112297}.

\bibitem{ISSProject}
\bibinfo{author}{\bibfnamefont{P.}~\bibnamefont{Huber}},
  \bibinfo{author}{\bibfnamefont{M.}~\bibnamefont{Lindner}},
  \bibinfo{author}{\bibfnamefont{M.}~\bibnamefont{Rolinec}}, \bibnamefont{and}
  \bibinfo{author}{\bibfnamefont{W.}~\bibnamefont{Winter}}  (\bibinfo{year}{to
  appear}).

\bibitem{Donini:2005db}
\bibinfo{author}{\bibfnamefont{A.}~\bibnamefont{Donini}},
  \bibinfo{author}{\bibfnamefont{E.}~\bibnamefont{Fernandez-Martinez}},
  \bibinfo{author}{\bibfnamefont{D.}~\bibnamefont{Meloni}}, \bibnamefont{and}
  \bibinfo{author}{\bibfnamefont{S.}~\bibnamefont{Rigolin}}
  \eprint{hep-ph/0512038}.

\bibitem{Donini:2005gy}
\bibinfo{author}{\bibfnamefont{A.}~\bibnamefont{Donini}},
  \bibinfo{author}{\bibfnamefont{E.}~\bibnamefont{Fernandez-Martinez}},
  \bibnamefont{and} \bibinfo{author}{\bibfnamefont{S.}~\bibnamefont{Rigolin}}
  \eprint{hep-ph/0509349}.

\bibitem{Barger:2004db}
\bibinfo{author}{\bibfnamefont{V.}~\bibnamefont{Barger}},
  \bibinfo{author}{\bibfnamefont{S.}~\bibnamefont{Geer}}, \bibnamefont{and}
  \bibinfo{author}{\bibfnamefont{K.}~\bibnamefont{Whisnant}},
  \bibinfo{journal}{New J. Phys.} \textbf{\bibinfo{volume}{6}},
  \bibinfo{pages}{135} (\bibinfo{year}{2004}), \eprint{hep-ph/0407140}.

\bibitem{Farzan:2002ct}
\bibinfo{author}{\bibfnamefont{Y.}~\bibnamefont{Farzan}} \bibnamefont{and}
  \bibinfo{author}{\bibfnamefont{A.~Y.} \bibnamefont{Smirnov}},
  \bibinfo{journal}{Phys. Rev.} \textbf{\bibinfo{volume}{D65}},
  \bibinfo{pages}{113001} (\bibinfo{year}{2002}), \eprint{hep-ph/0201105}.

\bibitem{Zhang:2004hf}
\bibinfo{author}{\bibfnamefont{H.}~\bibnamefont{Zhang}} \bibnamefont{and}
  \bibinfo{author}{\bibfnamefont{Z.-z.} \bibnamefont{Xing}},
  \bibinfo{journal}{Eur. Phys. J.} \textbf{\bibinfo{volume}{C41}},
  \bibinfo{pages}{143} (\bibinfo{year}{2005}), \eprint{hep-ph/0411183}.

\bibitem{Xing:2005gk}
\bibinfo{author}{\bibfnamefont{Z.-z.} \bibnamefont{Xing}} \bibnamefont{and}
  \bibinfo{author}{\bibfnamefont{H.}~\bibnamefont{Zhang}},
  \bibinfo{journal}{Phys. Lett.} \textbf{\bibinfo{volume}{B618}},
  \bibinfo{pages}{131} (\bibinfo{year}{2005}), \eprint{hep-ph/0503118}.

\bibitem{Blennow:2005yk}
\bibinfo{author}{\bibfnamefont{M.}~\bibnamefont{Blennow}},
  \bibinfo{author}{\bibfnamefont{T.}~\bibnamefont{Ohlsson}}, \bibnamefont{and}
  \bibinfo{author}{\bibfnamefont{W.}~\bibnamefont{Winter}},
  \bibinfo{journal}{JHEP} \textbf{\bibinfo{volume}{06}}, \bibinfo{pages}{049}
  (\bibinfo{year}{2005}), \eprint{hep-ph/0502147}.

\bibitem{Blennow:2005qj}
\bibinfo{author}{\bibfnamefont{M.}~\bibnamefont{Blennow}},
  \bibinfo{author}{\bibfnamefont{T.}~\bibnamefont{Ohlsson}}, \bibnamefont{and}
  \bibinfo{author}{\bibfnamefont{W.}~\bibnamefont{Winter}}
  \eprint{hep-ph/0508175}.

\bibitem{Huber:2002bi}
\bibinfo{author}{\bibfnamefont{P.}~\bibnamefont{Huber}},
  \bibinfo{author}{\bibfnamefont{T.}~\bibnamefont{Schwetz}}, \bibnamefont{and}
  \bibinfo{author}{\bibfnamefont{J.~W.~F.} \bibnamefont{Valle}},
  \bibinfo{journal}{Phys. Rev.} \textbf{\bibinfo{volume}{D66}},
  \bibinfo{pages}{013006} (\bibinfo{year}{2002}), \eprint{hep-ph/0202048}.

\bibitem{Winter:2005we}
\bibinfo{author}{\bibfnamefont{W.}~\bibnamefont{Winter}},
  \bibinfo{journal}{Phys. Rev.} \textbf{\bibinfo{volume}{D72}},
  \bibinfo{pages}{037302} (\bibinfo{year}{2005}), \eprint{hep-ph/0502097}.

\bibitem{Winter:2005ud}
\bibinfo{author}{\bibfnamefont{W.}~\bibnamefont{Winter}}
  \bibinfo{note}{Proceedings of {NuFact 05, June 21-26, 2005, Frascati,
  Italy}}, \eprint{hep-ph/0510025}.

\bibitem{Winter:2006vg}
\bibinfo{author}{\bibfnamefont{W.}~\bibnamefont{Winter}}
  \bibinfo{note}{Proceedings of {Neutrino sciences 2005: Neutrino geophysics,
  December 14-16, 2005, Honolulu, USA}}, \eprint{physics/0602049}.

\bibitem{Autin:2002ms}
\bibinfo{author}{\bibfnamefont{B.}~\bibnamefont{Autin}} \emph{et~al.},
  \bibinfo{journal}{J. Phys.} \textbf{\bibinfo{volume}{G29}},
  \bibinfo{pages}{1785} (\bibinfo{year}{2003}), \eprint{physics/0306106}.

\bibitem{Mezzetto:2003ub}
\bibinfo{author}{\bibfnamefont{M.}~\bibnamefont{Mezzetto}},
  \bibinfo{journal}{J. Phys.} \textbf{\bibinfo{volume}{G29}},
  \bibinfo{pages}{1771} (\bibinfo{year}{2003}), \eprint{hep-ex/0302007}.

\bibitem{Bouchez:2003fy}
\bibinfo{author}{\bibfnamefont{J.}~\bibnamefont{Bouchez}},
  \bibinfo{author}{\bibfnamefont{M.}~\bibnamefont{Lindroos}}, \bibnamefont{and}
  \bibinfo{author}{\bibfnamefont{M.}~\bibnamefont{Mezzetto}},
  \bibinfo{journal}{AIP Conf. Proc.} \textbf{\bibinfo{volume}{721}},
  \bibinfo{pages}{37} (\bibinfo{year}{2004}), \eprint{hep-ex/0310059}.

\bibitem{Lindroos:2003kp}
\bibinfo{author}{\bibfnamefont{M.}~\bibnamefont{Lindroos}}
  \eprint{physics/0312042}.

\bibitem{Lindroos:2004sa}
\bibinfo{author}{\bibfnamefont{M.}~\bibnamefont{Lindroos}} \bibnamefont{and}
  \bibinfo{author}{\bibfnamefont{C.}~\bibnamefont{Volpe}},
  \bibinfo{journal}{Nucl. Phys. News} \textbf{\bibinfo{volume}{14}},
  \bibinfo{pages}{15} (\bibinfo{year}{2004}).

\bibitem{Couce}
\bibinfo{author}{\bibfnamefont{E.}~\bibnamefont{Couce}},
  \emph{\bibinfo{title}{Comparison of facilities using a {Water Cherenkov}
  detector}}, \bibinfo{note}{{Talk given at the 2nd International Scoping study
  meeting of Neutrino-factory and Superbeam Facility, January 24, 2006, KEK,
  Tsukuba, Japan}}.

\bibitem{Volpe:2003fi}
\bibinfo{author}{\bibfnamefont{C.}~\bibnamefont{Volpe}}, \bibinfo{journal}{J.
  Phys.} \textbf{\bibinfo{volume}{G30}}, \bibinfo{pages}{L1}
  (\bibinfo{year}{2004}), \eprint{hep-ph/0303222}.

\bibitem{McLaughlin:2003yg}
\bibinfo{author}{\bibfnamefont{G.~C.} \bibnamefont{McLaughlin}}
  \bibnamefont{and} \bibinfo{author}{\bibfnamefont{C.}~\bibnamefont{Volpe}},
  \bibinfo{journal}{Phys. Lett.} \textbf{\bibinfo{volume}{B591}},
  \bibinfo{pages}{229} (\bibinfo{year}{2004}), \eprint{hep-ph/0312156}.

\bibitem{Serreau:2004kx}
\bibinfo{author}{\bibfnamefont{J.}~\bibnamefont{Serreau}} \bibnamefont{and}
  \bibinfo{author}{\bibfnamefont{C.}~\bibnamefont{Volpe}},
  \bibinfo{journal}{Phys. Rev.} \textbf{\bibinfo{volume}{C70}},
  \bibinfo{pages}{055502} (\bibinfo{year}{2004}), \eprint{hep-ph/0403293}.

\bibitem{Burguet-Castell:2005pa}
\bibinfo{author}{\bibfnamefont{J.}~\bibnamefont{Burguet-Castell}},
  \bibinfo{author}{\bibfnamefont{D.}~\bibnamefont{Casper}},
  \bibinfo{author}{\bibfnamefont{E.}~\bibnamefont{Couce}},
  \bibinfo{author}{\bibfnamefont{J.~J.} \bibnamefont{Gomez-Cadenas}},
  \bibnamefont{and}
  \bibinfo{author}{\bibfnamefont{P.}~\bibnamefont{Hernandez}},
  \bibinfo{journal}{Nucl. Phys.} \textbf{\bibinfo{volume}{B725}},
  \bibinfo{pages}{306} (\bibinfo{year}{2005}), \eprint{hep-ph/0503021}.

\bibitem{Huber:2005jk}
\bibinfo{author}{\bibfnamefont{P.}~\bibnamefont{Huber}},
  \bibinfo{author}{\bibfnamefont{M.}~\bibnamefont{Lindner}},
  \bibinfo{author}{\bibfnamefont{M.}~\bibnamefont{Rolinec}}, \bibnamefont{and}
  \bibinfo{author}{\bibfnamefont{W.}~\bibnamefont{Winter}},
  \bibinfo{journal}{Phys. Rev.} \textbf{\bibinfo{volume}{D}} (\bibinfo{year}{to
  appear}), \eprint{hep-ph/0506237}.

\bibitem{Agarwalla:2005we}
\bibinfo{author}{\bibfnamefont{S.~K.} \bibnamefont{Agarwalla}},
  \bibinfo{author}{\bibfnamefont{A.}~\bibnamefont{Raychaudhuri}},
  \bibnamefont{and} \bibinfo{author}{\bibfnamefont{A.}~\bibnamefont{Samanta}},
  \bibinfo{journal}{Phys. Lett.} \textbf{\bibinfo{volume}{B629}},
  \bibinfo{pages}{33} (\bibinfo{year}{2005}), \eprint{hep-ph/0505015}.

\bibitem{Donini:2004hu}
\bibinfo{author}{\bibfnamefont{A.}~\bibnamefont{Donini}},
  \bibinfo{author}{\bibfnamefont{E.}~\bibnamefont{Fernandez-Martinez}},
  \bibinfo{author}{\bibfnamefont{P.}~\bibnamefont{Migliozzi}},
  \bibinfo{author}{\bibfnamefont{S.}~\bibnamefont{Rigolin}}, \bibnamefont{and}
  \bibinfo{author}{\bibfnamefont{L.}~\bibnamefont{Scotto~Lavina}},
  \bibinfo{journal}{Nucl. Phys.} \textbf{\bibinfo{volume}{B710}},
  \bibinfo{pages}{402} (\bibinfo{year}{2005}), \eprint{hep-ph/0406132}.

\bibitem{Donini:2004iv}
\bibinfo{author}{\bibfnamefont{A.}~\bibnamefont{Donini}},
  \bibinfo{author}{\bibfnamefont{E.}~\bibnamefont{Fernandez-Martinez}},
  \bibnamefont{and} \bibinfo{author}{\bibfnamefont{S.}~\bibnamefont{Rigolin}},
  \bibinfo{journal}{Phys. Lett.} \textbf{\bibinfo{volume}{B621}},
  \bibinfo{pages}{276} (\bibinfo{year}{2005}), \eprint{hep-ph/0411402}.

\bibitem{EURISOL}
\emph{\bibinfo{title}{Eurisol design study}}, \bibinfo{note}{{\tt
  http://www.eurisol.org}}.

\end{thebibliography}
\end{document}